\RequirePackage{amsmath}
\documentclass[runningheads]{llncs}
\usepackage[T1]{fontenc}
\usepackage[table,xcdraw]{xcolor}

\usepackage{relsize}
\usepackage{scalerel}
\usepackage{xspace}
\usepackage{cancel}
\usepackage{amssymb}
\usepackage{bbding}
\usepackage{booktabs}
\usepackage{AdditionalCommands}
\usepackage{multirow}
\usepackage{subcaption}
\definecolor{orcid-green}{RGB}{166, 206, 57}

\DeclareRobustCommand\orcidlink[1]{%
  \texorpdfstring{\href{https://orcid.org/#1}{{\raisebox{-.5ex}{\includegraphics[height=1.1em]{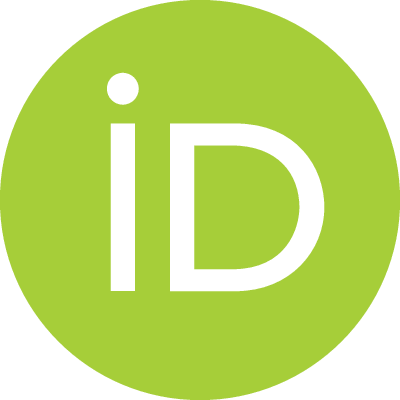}}}}}{https://orcid.org/#1}%
}

\PassOptionsToPackage{hyphens}{url}
\usepackage{hyperref}

\renewcommand{\footnotesize}{\scriptsize}

\begin{document}
\title{Evaluation of Study Plans using Partial Orders}
\author{Christian Rennert\href{mailto:rennert@pads.rwth-aachen.de}{\Envelope} \orcidlink{0000-0003-4614-6171}\and
Mahsa Pourbafrani \orcidlink{0000-0002-7883-1627} \and
Wil van der Aalst \orcidlink{0000-0002-0955-6940}}

\authorrunning{C. Rennert et al.}

\institute{Chair of Process and Data Science (PADS), RWTH Aachen University, Germany
\email{\{rennert,mahsa.pourbafrani,wvdaalst\}@pads.rwth-aachen.de}}

\maketitle              % typeset the header of the contribution
\begin{abstract}
In higher education, data is collected that indicate the term(s) that a course is taken and when it is passed.
Often, study plans propose a suggested course order to students.
Study planners can adjust these based on detected deviations between the proposed and actual order of the courses being taken.
In this work, we detect deviations by combining (1)~the deviation between the proposed and actual course order with (2)~the temporal difference between the expected and actual course-taking term(s).
Partially ordered alignments identify the deviations between the proposed and actual order.
We compute a partial order alignment by modeling a study plan as a process model and a student's course-taking behavior as a partial order.
Using partial orders in such use cases allows one to relax the constraints of strictly ordered traces.
This makes our approach less prone to the order in which courses are offered.
Further, when modeling course-taking behavior as partial orders, we propose distinguishing intended course-taking behavior from actual course-passing behavior of students by including either all terms in which a course is attempted or only the term that a course is passed, respectively.
This provides more perspectives when comparing the proposed and actual course-taking behavior.
The proposed deviation measuring approach is evaluated on real-life data from RWTH Aachen University.

\end{abstract}
\section{Introduction}\label{sec:intro}
In Germany, universities must provide students with the knowledge, skills, and methods required for their subjects.\footnote{
In Germany, states regulate higher education.
For example, North Rhine-Westphalia defines the purpose of study plans in \textsection 58 (1) Hochschulgesetz (HG) NRW.
}
Furthermore, regulations state that (1)~every degree program must offer a study plan completable within the degree's standard duration, and (2)~universities must create individual study plans for students if required.\footnote{According to \textsection 58 (3) Hochschulgesetz (HG) NRW.}
In this work, a study plan comprises several courses whose completion is assessed by examination activities.
Here, only the passing of final exams is considered as such activities, although this is not a general restriction.

Across all study programs offered in 2022 in Germany, nearly 247,000 students received their bachelor's degree\footnote{\url{https://www.statista.com/statistics/584454/bachelor-and-master-degrees-number-universities-germany/}, 2024, last access 2024-07-05}.
First-time graduates received their degree on average after around 4 years of studies\footnote{\url{https://www.statista.com/statistics/584277/average-study-duration-graduates-germany/}, 2023, last access 2024-07-05.}.
However, the standard period of study for bachelor's degree courses in Germany is three years.
This indicates an average extension of studies by around one year in 2022.

In \cite{penthin2017grunde}, potential reasons for an individual to extend the duration of their study are categorized.
In detail, they identify (1) study conditions, (2) individual characteristics and entry requirements, and (3) personal living conditions and context factors as categories of reasons.
Although there are many reasons for an extended study duration, if possible, study planners should implement measurements to avoid an extended study duration.
Therefore, a systematic analysis of student data and study plans helps identify and improve study plan shortcomings such that study conditions are improved.

\begin{figure}[t]
    \centering
    \includegraphics[width=\linewidth]{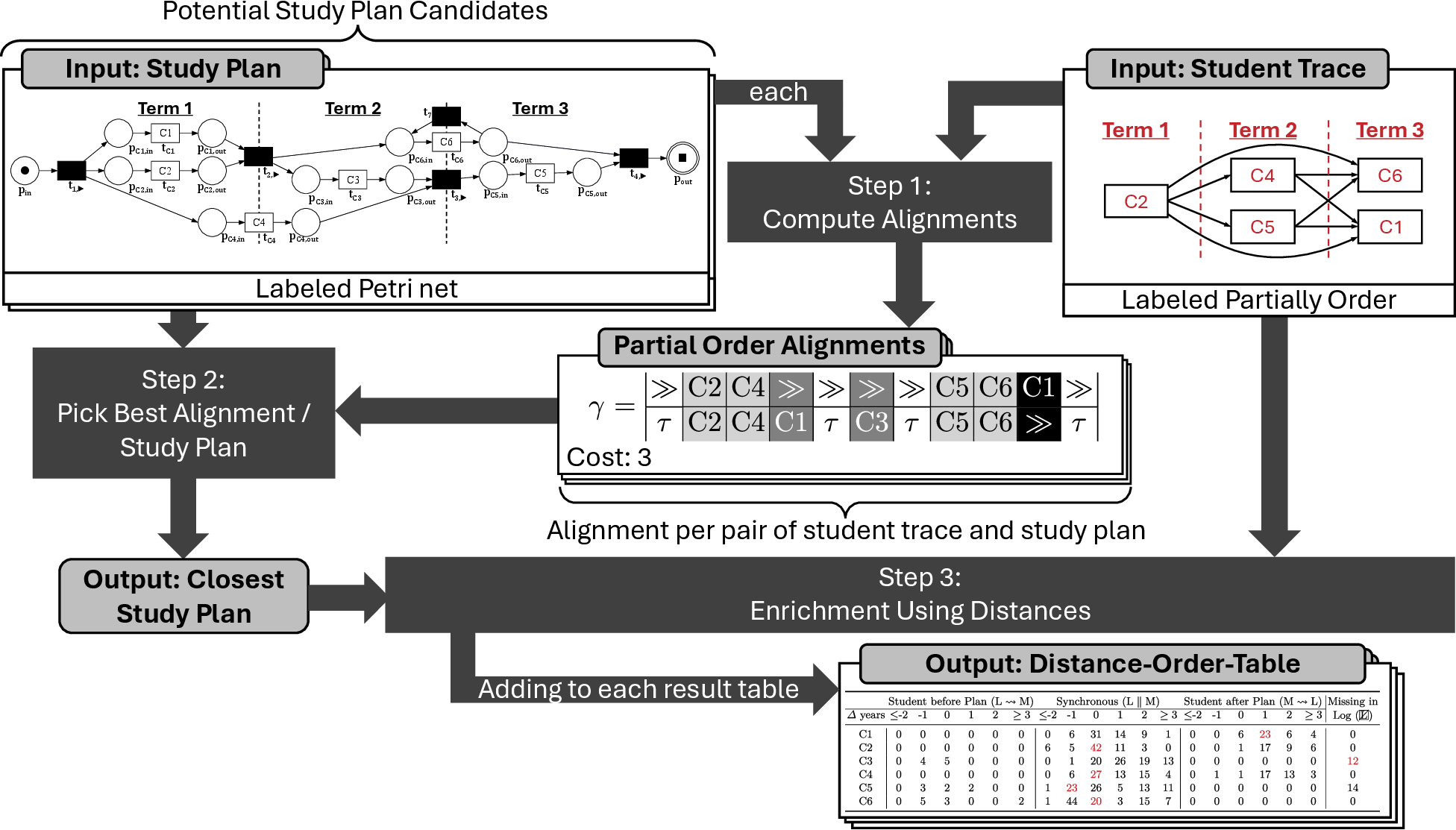}
    \caption{Framework for the evaluation of study plans using partial orders.}
    \label{fig:overview}
\end{figure}

One approach is to use process mining to understand students' interactions with their study plan.
In detail, we model study plans as process models and store exam data in event logs to allow for systematic analysis.

The complete approach presented in this work is the following.
Workflow nets are used to model all study plans for a given time period, as there may be multiple study plans available due to changes applied to a degree over time.
Partial orders are used to model the student's course-taking order.
Both are compared using partial order alignments.
The alignment with the lowest cost is then identified.
For its study plan and all its courses, we aggregate counts for each combination of (1) the course's relative position between the log and model parts of the alignment, indicating whether a course is moved forward, backward, or if it is in the right position in the study plan, and (2) the absolute difference between the expected and actual terms that the course is taken.
This procedure is visualized in Figure~\ref{fig:overview}.

The remainder of this paper is structured as follows.
Section \ref{sec:related} details the related work, Section \ref{sec:prelims} gives the preliminaries for this work, Section \ref{sec:approach} describes the approach introduced in the paper, Section \ref{sec:eval} evaluates the results obtained by applying the approach to a real-life dataset, and Section \ref{sec:conclusion} concludes the paper.

\section{Related Work}\label{sec:related}
Process mining in the context of education, also known as Educational Process Mining~(EPM), is the subject of several papers.
An extensive survey by Bogarín et al.~\cite{DBLP:journals/widm/BogarinCR18} discusses educational process mining as an educational data mining~(EDM) technique.
Here, the authors show that there are some works using conformance checking techniques to identify the agreement between students' exam histories as event traces and their study plans as process models.
Alternatively, the authors in \cite{DBLP:journals/widm/BogarinCR18} also discuss that conformance checking can be applied using rule modeling for expected student behavior, which is another conformance checking technique not investigated in this paper.
For more details on EPM, we refer to the survey in \cite{DBLP:journals/widm/BogarinCR18}.
We limit our scope to works that target the course-taking process of students and their interactions with the study plan.

Hobeck et al.\@~\cite{DBLP:conf/icpm/HobeckPW22} apply process mining and in particular the PM$^2$ methodology on exam data from the university information system of the TU Berlin.
For a group of students, the authors investigate the question of whether students follow their study plan by comparing its conformance with the full student's exam history.
In the work of Bendatu and Yahya~\cite{bendatu_sequence_2015}, they also compare exam histories with a single study plan by computing the temporal distance between the expected and actual point in time when an exam is taken.
We use the general idea and extend the study plan to allow for several expected terms.

In~\cite{DBLP:journals/eswa/DiamantiniGMPZ24}, Diamantini et al.\@ model study plans as blocks of exams to be taken to compare the ordering of exams in the student's exam history and in the study plan.
They use their approach to understand bottlenecks and to make a distinction between successful and non-successful students.
Similar to abstracting study plans into course blocks, some works model student exam histories as blocks, using a set abstraction per term.
In \cite{DBLP:journals/access/PriyambadaEYU21}, Priyambada et al. use this approach to compare and cluster students by detecting course patterns for the relative term course sets.
In \cite{DBLP:conf/icpm/RafieiBPPHLA23}, Rafiei et al.\@ exploit atomic ordering and educational KPI information in exam data to derive labels for rule extraction with the aim to assist study planners.
In their work, they derived partial orders by considering DFGs to directly represent a partial order.  

Usually in process mining, the ordering based on timestamps between events and a suggested ordering of activities in process models can be compared using alignments as first introduced by Adriansyah~\cite{AlignmentsAdriansyah}.
Later, in \cite{xixi_partial_conformance_2015}, Lu et al.\@ extended alignments to also be applicable for partial orders.
Nevertheless, so far no existing work incorporates partial order alignments in its research or case study.
Therefore, this is among the first works to incorporate partial orders.

In the following section, we discuss the preliminaries of this work.
\section{Preliminaries}\label{sec:prelims}
\noindent
\begin{minipage}[t]{0.6\linewidth}
    \paragraph{Basic Notations.}
    A \emph{multiset} generalizes the concept of a set and allows for multiple occurrences of the same element, e.g., \amulset{\elemx,\elemy,\elemy,\elemx,\elemx,\elemz} = \amulset{\elemx^3,\elemy^2,\elemz}.
    The set of all multisets over a set~\setX is~\multisetset{\setX}.
    Similarly to a multiset, a \emph{sequence} can contain multiple occurrences of the same element.
    However, within a sequence, all elements are \emph{totally ordered}, e.g., $\trace = \Trace{\elemx,\elemy,\elemy,\elemx,\elemz}$ is a sequence over the set~$\setX = \aset{\elemx,\elemy,\elemz}$ and thus it is in its \emph{Kleene closure} $\kleeneClosure{\setX}$, i.e., $\trace \in \kleeneClosure{\setX}$.
    Given a sequence $\trace$ and a value $i \in \naturals$ with $1 \leq i \leq \cardinality{\trace}$, we denote by $\trace(i)$ the element at the $i$-th position of the sequence $\trace$, e.g., for \makebox[\linewidth][s]{$\trace = \Trace{\elemx,\elemy,\elemz}$ it holds that $\trace(3) = \elemz$.
    Given an}
\end{minipage}\hfill
\begin{minipage}[t]{0.37\linewidth}
    \centering
    \captionsetup{type=table}
    \captionof{table}{Example educational event log $\log[1]$.}
    \resizebox{\linewidth}{!}{
    \begin{minipage}{\linewidth}
        \centering
        \label{tab:edu_event_log}
        \begin{tabular}{ccccccccc}
            \toprule
            Student                 & Course                  & Term                   & State                       \\ \midrule
            \color[HTML]{C1272D} S1 & \color[HTML]{C1272D} C2 & \color[HTML]{C1272D} 1 & \color[HTML]{C1272D} 1 \\
            S2 & C1 & 2 & 1 \\
            \color[HTML]{C1272D} S1 & \color[HTML]{C1272D} C4 & \color[HTML]{C1272D} 2 & \color[HTML]{C1272D} 1 \\
            S2 & C4 & 2 & 0 \\
            S2 & C4 & 2 & 1 \\
            \color[HTML]{C1272D} S1 & \color[HTML]{C1272D} C5 & \color[HTML]{C1272D} 2 & \color[HTML]{C1272D} 1 \\
            \color[HTML]{C1272D} S1 & \color[HTML]{C1272D} C6 & \color[HTML]{C1272D} 3 & \color[HTML]{C1272D} 1 \\
            \color[HTML]{C1272D} S1 & \color[HTML]{C1272D} C1 & \color[HTML]{C1272D} 3 & \color[HTML]{C1272D} 1 \\ \bottomrule
        \end{tabular}
    \end{minipage}
    }
\end{minipage}
\noindent
$n$-tuple $(\elema[1], \elema[2], \dots, \elema[n])$, we define $\sel{i}$ to be the selector function that returns the i-th entry, e.g., $\sel{2}((a,b,c)) = b$.
We define that applying a single element function $f: X \rightarrow Y$ to a sequence $\trace \in \kleeneClosure{X}$, applies the function $f$ to each element in the sequence, i.e., $f(\trace) = \Trace{f(x) \mid x \in \trace}$.
The domain of a function $f$ is described by $dom(f)$, e.g., $dom(f) = \aset{1,2,3}$ for $f(1) = 4$, $f(2) = 5$, and $f(3) = 3$. 

\paragraph{Educational Event Log.}
In process mining, \emph{event data} are used to observe and keep track of \emph{behavior}.
In our use case, an event exists for each exam try of a student.
Therefore, we first introduce the universe of student matriculation numbers \universeofMatriculationNumbers and the universe of course IDs \universeofCourseIDs.
An event $\event \in \universeofMatriculationNumbers \times \universeofCourseIDs \times \naturals \times \aset{0, 1}$ is a quadruple of a student's matriculation number, a course ID of a taken course, a natural representing the term in which the event occurred relative to the start of the student's studies, and a state indicating whether the exam attempt is passed (1) or failed (0).
We may also refer to the matriculation number as student ID.
The universe of events is denoted by \universeofevents.
An example of an event is $\event[1] = (\text{S2}, \text{C4}, 2, 1) \in \universeofevents$ which shows that the course C4 is passed by the student S2 in its second term of studies.
An event log $\log \in \multisetset{\universeofevents}$ is a multiset of events, as there can be several exam attempts within a term.

Given an event $\event$, for better readability, we define $\sel{mat}(\event) = \sel{1}(\event)$, $\sel{cid}(\event) = \sel{2}(\event)$, $\sel{term}(\event) = \sel{3}(\event)$, and $\sel{pass}(\event) = \sel{4}(\event)$, e.g., $\sel{cid}(\event[1]) = \text{C4}$.
In this work, we restrict each educational event log $\log$ to only contain courses that are eventually passed, i.e., $\forall_{\event \in \log}\exists_{\event' \in \log} \colon ((\sel{cid}(\event) {=} \sel{cid}(\event')) \wedge (\sel{pass}(\event') {=} 1))$. 
An example of an educational event log $\log[1]$ is given in Table~\ref{tab:edu_event_log} where $\event[1] \in \log[1]$ holds.
\strut

\noindent
\begin{minipage}[b]{0.6\linewidth}
    \paragraph{Partial Orders.}
    We use \emph{partial orders} to model temporal relations between student events, these are used because multiple events an occur for the same term and we do not consider the order within a term.
    In detail, labeled partial orders are used, as we may have to model students who took several courses with the same course ID.
    This might be the case for a student that is \makebox[\linewidth][s]{required to take distinct seminars with the same}
\end{minipage}\hfill
\begin{minipage}[b]{0.37\linewidth}
    \centering
    \captionsetup{type=figure}
    \includegraphics[width=\linewidth]{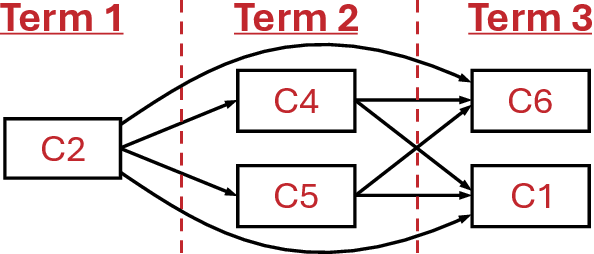}
    \captionof{figure}{{\color[HTML]{C1272D}Labeled partial order $lpo_{S1}$} of passed courses for the student with student ID {\color[HTML]{C1272D}S1}.}
    \label{fig:dag_e1}
\end{minipage}

\noindent
course ID.
A labeled partial order $lpo$ is a quadruple~$lpo = (\partialNodes,\partialRelation,\activities,\labelingfunc)$ where \partialNodes~is a set of nodes, \partialRelation~is a partial order relation, \activities~is a set of activities, and $\labelingfunc \colon \partialNodes \rightarrow \activities$ is a labeling function.
Figure~\ref{fig:dag_e1} shows the labeled partial order for the student with {\color[HTML]{C1272D}Student ID~S1} from event log $\log[1]$.
The next section describes how to derive labeled partial orders from an event log.
First, we introduce labeled Petri nets as process models so that we can model study plans.

\paragraph{Petri and Workflow Nets.}
A \emph{labeled Petri net} $\net$ is a quintuple $\net=(\places, \transitions, \arcs, \activities, \labelingfunc)$, where \places is a finite set of \emph{places}, \transitions is a finite set of \emph{transitions} such that $\places \cap \transitions = \emptyset$, ${\arcs \subseteq (\places \times \transitions) \cup (\transitions \times \places)}$ is a set of directed arcs, the so-called \emph{flow relation}, \activities is a set of activities, and $\labelingfunc \colon \transitions \rightarrow \activities \cup \aset{\tau}$ is a \emph{labeling function} with $\tau \not \in \activities$ being a reserved silent activity label corresponding to no activity being performed.

\begin{figure}[t]
    \centering
    \includegraphics[width=\linewidth]{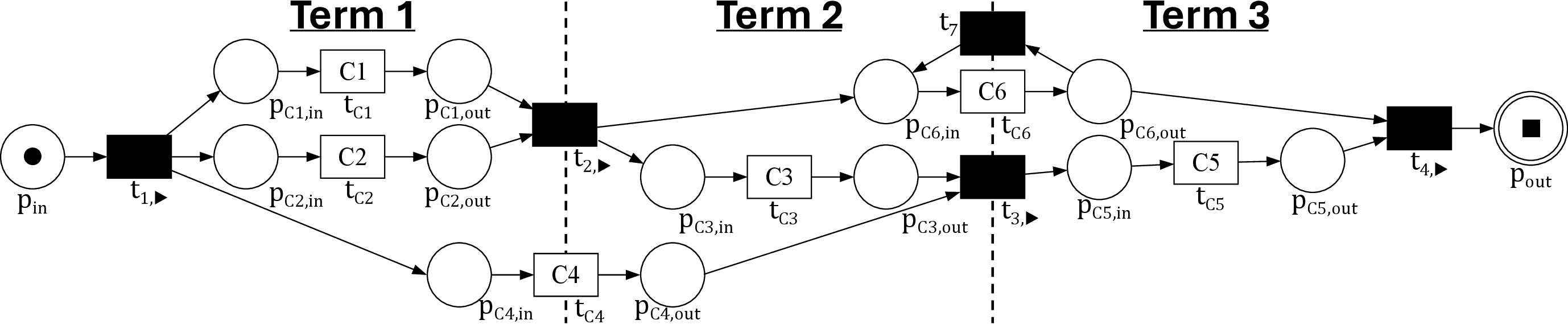}
    \caption{Example workflow net \net[1].}
    \label{fig:example-workflow-net}
\end{figure}

A \emph{workflow net} is a labeled Petri net $\net = (\places, \transitions, \arcs, \activities, \labelingfunc)$ for which there is exactly one input place $in \in \places$ such that $\aset{(\place, in) \in \arcs \mid \place \in \places} = \emptyset$, one output place $out \in \places$ such that $\aset{(out, \place) \in \arcs \mid \place \in \places} = \emptyset$, and for which modifying it to be short-circuited makes it strongly connected.
An example workflow net is shown in Figure~\ref{fig:example-workflow-net}.
For more details on Petri nets as process models and their semantics, we refer to \cite{DBLP:books/sp/Aalst16}.
Workflow nets and labeled partial orders describing observed behavior can be checked for conformance using alignments.

\paragraph{Alignments.}
A partial order alignment $\alignment(lpo, \net)$ relates the behavior described by a labeled partial order $lpo$ with the behavior of a workflow net $\net$.
Therefore, an alignment consists of several moves that distinguish whether the behaviors agree and are in synchronization or whether they disagree and deviate.
In detail, we distinguish the following four move types.
(1) \emph{Synchronous moves} (\MyColorBox[lightgray]{lightgray}) indicate no deviation,
(2) \emph{log moves} (\MyColorBox[black]{\color{white}black}) indicate a deviation such that a move was taken on the log side but not on the model side,
(3) \emph{model moves} (\MyColorBox[gray]{\color{white}gray}) indicate a deviation such that a move was taken on the log side but not on the model side,
and (4) \emph{invisible model moves} indicated by $\tau$ specify that a not observable model move is taken and that a misalignment on the log side is not required.
In a deviation, the misalignment is denoted by ``$\nomove$'', to indicate that either the log or the model does not have any matching behavior.

An alignment that best matches the model and partial order is the so-called \emph{optimal alignment} because it has a minimal number of mismatches.
In Table~\ref{table:example_alignment}, an optimal alignment $\alignment(lpo_{S1}, \net[1]) = \Trace{(\nomove, \tau), (\text{C2}, \text{C2}), \dots, (\nomove, \tau)}$ between the labeled partial order $lpo_{S1}$ and the workflow net $\net{1}$ is given.
The optimal alignment $\alignment(lpo_{S1}, \net[1])$ has three misalignments, i.e., one log and two model moves.
The upper row indicates the log behavior, while the lower row indicates the model behavior.
For an alignment \alignment, we define $\alignmentLog = \sel{1}(\alignment)$ and $\alignmentModel = \sel{2}(\alignment)$, e.g., $\alignmentLog(lpo_{S1}, \net[1]) = \Trace{\nomove, \text{C2}, \text{C4}, \nomove, \nomove, \nomove, \nomove, \text{C5}, \text{C6}, \text{C1}, \nomove}$
and $\alignmentModel(lpo_{S1}, \net[1]) = \Trace{\tau, \text{C2}, \text{C4}, \text{C1}, \tau, \text{C3}, \tau, \text{C5}, \text{C6}, \nomove, \tau}$ for the behavior on the log and model side, respectively.
We can now consider the impact of using labeled partial orders instead of ordering courses within a term.
Consider a labeled partial order $lpo_{S1}'$ that requires $C4$ and $C6$ to always occur before $C5$ and $C1$, respectively, although events $(\text{S1}, \text{C4}, 2, 1)$ together with $(\text{S1}, \text{C5}, 2, 1)$ and $(\text{S1}, \text{C6}, 3, 1)$ together with $(\text{S1}, \text{C1}, 3, 1)$ each have the same term, which otherwise indicates no strict ordering.
Computing an optimal alignment $\alignment(lpo_{S1}', \net[1])$, as shown in Table \ref{table:example_alignment2}, between $lpo_{S1}'$ and $\net[1]$ results in a worse alignment than the alignment $\alignment(lpo_{S1}, \net[1])$, since its cost is higher by a value of two.
Therefore, abstracting course orderings within the same term affects alignment computation.
For details on the computation of partial order alignments, we refer to \cite{xixi_partial_conformance_2015}.

\noindent
\begin{minipage}{0.48\linewidth}
    \resizebox{\textwidth}{!}{
    \begin{minipage}{1.2\linewidth}
        \centering
        \captionof{table}{An optimal alignment $\alignment(lpo_{S1}, \net[1])$ for $lpo_{S1}$ and $\net[1]$.}
        \label{table:example_alignment}
        \begin{tabular}{|c|
        >{\columncolor[HTML]{D3D3D3}}c |
        >{\columncolor[HTML]{D3D3D3}}c |
        >{\columncolor[HTML]{808080}}c |c|
        >{\columncolor[HTML]{808080}}c |c|
        >{\columncolor[HTML]{D3D3D3}}c |
        >{\columncolor[HTML]{D3D3D3}}c |
        >{\columncolor[HTML]{000000}}c |c|}
            \nomove                  & C2                   & C4                   & \color[HTML]{FFFFFF} \nomove              & \nomove                  & \color[HTML]{FFFFFF} \nomove              & \nomove                  & C5                    & C6                   & \color[HTML]{FFFFFF} C1      & \nomove                   \\ \hline
            $\tau$ & C2 & C4 & \color[HTML]{FFFFFF} C1 & $\tau$ & \color[HTML]{FFFFFF} C3 & $\tau$ & C5 & C6 & \color[HTML]{FFFFFF} \nomove & $\tau$ 
        \end{tabular}
    \end{minipage}
    }
\end{minipage}\hfill
\begin{minipage}{0.48\linewidth}
    \resizebox{\textwidth}{!}{
    \begin{minipage}{1.2\linewidth}
        \centering
        \captionof{table}{An optimal alignment $\alignment(lpo_{S1}', \net[1])$ for $lpo_{S1}'$ and $\net[1]$.}
        \label{table:example_alignment2}
        \begin{tabular}{|c|
        >{\columncolor[HTML]{D3D3D3}}c |
        >{\columncolor[HTML]{000000}}c |
        >{\columncolor[HTML]{D3D3D3}}c |
        >{\columncolor[HTML]{808080}}c |c|
        >{\columncolor[HTML]{808080}}c |c|
        >{\columncolor[HTML]{808080}}c |
        >{\columncolor[HTML]{D3D3D3}}c |
        >{\columncolor[HTML]{000000}}c |c|}
            \nomove & C2 & \color[HTML]{FFFFFF} C5 & C4 & \color[HTML]{FFFFFF} \nomove & \nomove & \color[HTML]{FFFFFF} \nomove & \nomove & \color[HTML]{FFFFFF}\nomove & C6 & \color[HTML]{FFFFFF} C1 & \nomove \\ \hline
            $\tau$ & C2 & \color[HTML]{FFFFFF} \nomove & C4 & \color[HTML]{FFFFFF} C1 & $\tau$ & \color[HTML]{FFFFFF} C3 & $\tau$ & \color[HTML]{FFFFFF}C5 & C6 & \color[HTML]{FFFFFF} \nomove & $\tau$ 
        \end{tabular}
    \end{minipage}
    }
\end{minipage}
\section{Approach}\label{sec:approach}
\paragraph{Study Plan.}
For each degree program, study planners provide a study plan.
It organizes the degree program into terms, stating when each course should be taken.
In addition, it differentiates between mandatory courses, required to complete a degree program, and elective courses, of which only a subset must be taken.
We take electives into account by mapping all elective course IDs to a reserved placeholder elective course ID $C_{el} \not \in \universeofCourseIDs$.
A study plan process model should allow for the elective course ID $C_{el}$ to reoccur multiple times, modeling an unknown number of electives to be taken by an individual student.
Formally, we define a study plan $SP$ as $SP \colon (\universeofCourseIDs \cup \aset{C_{el}}) \rightarrow \naturals \times \naturals$ mapping a course ID $cid \in (\universeofCourseIDs \cup \aset{C_{el}})$ to its first expected term $T_{start}$ and its last expected term~$T_{end}$, i.e., $SP(cid) = (T_{start}, T_{end})$.
Next, we create a workflow net from a study plan to compute labeled partial orders.

\begin{figure}[t]
    \centering
    \includegraphics[width=\linewidth]{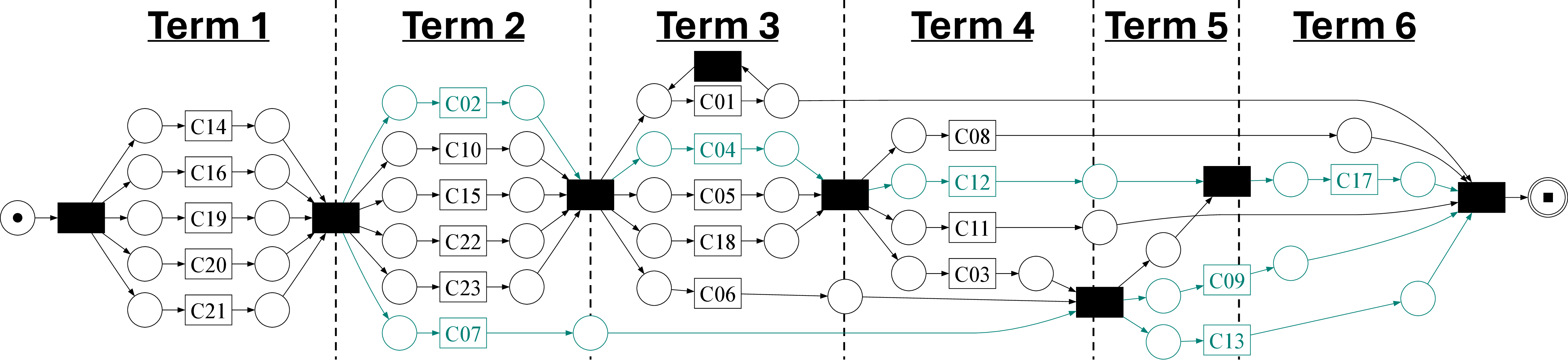}
    \caption{Workflow net $\net[10]$ from the study plan $SP_{10}$.}
    \label{fig:sp_petri_net_2010}
    
    \includegraphics[width=\linewidth]{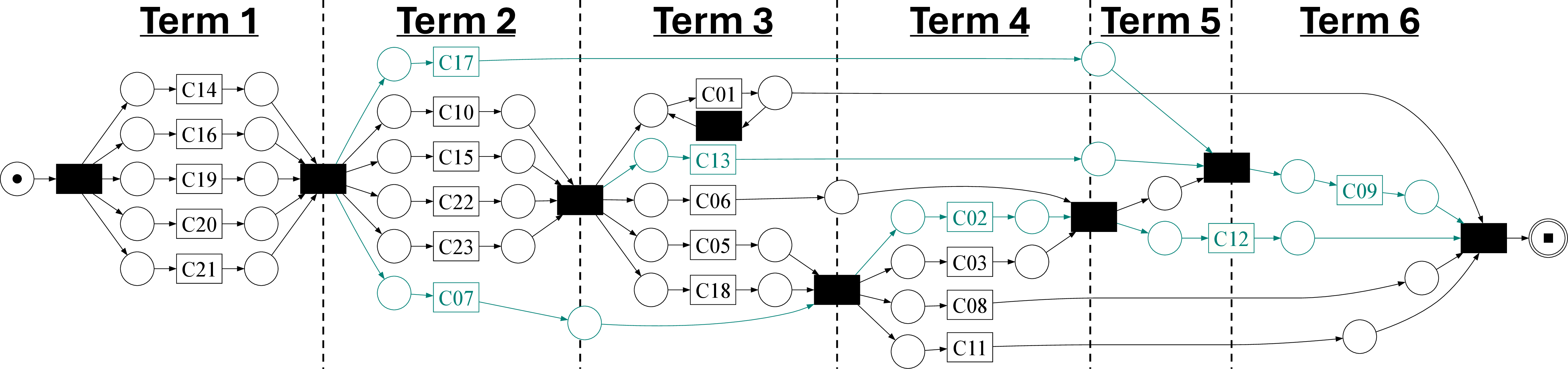}
    \caption{Workflow net $\net[18]$ from the study plan $SP_{18}$.}
    \label{fig:sp_petri_net_2018}
\end{figure}

\paragraph{Study Plan to Workflow Net.}
A workflow net $\net[SP] = (\places, \transitions, \arcs, \activities, \labelingfunc)$ is constructed from a study plan $SP$ as follows.
\textbf{Introducing term start and end synchronizing silent transitions:}
For each term $term \in \naturals$ with $1 \leq term \leq k$ contained in the study plan $SP$, with $k = max(\aset{\sel{2}(SP(cid)) \mid cid \in dom(SP)})$ being the highest term where $max$ is the maximum function, we add a silent transition $\transitionbracket[term, \sactivity]$ with $\labelingfunc(\transitionbracket[term, \sactivity]) = \tau$ starting each term.
We define that each transition $\transitionbracket[term, \sactivity]$ starting a term $term$ is the transition $\transitionbracket[(term - 1), \sactivity]$ that closes the previous term, i.e., $\forall_{term \in \range{2,k}} \colon \transitionbracket[(term - 1), \eactivity] = \transitionbracket[term, \sactivity]$.
Since there is a missing silent transition that closes the term $k$, we add $\transitionbracket[k, \eactivity]$ with $\labelingfunc(\transitionbracket[k, \eactivity]) = \tau$ to the set of transitions~$\transitions$.
We connect the transition that starts the first term with the input place $in \in \places$ and the transition that ends the last term with the output place $out \in \places$, i.e., $\aset{(in, \transitionbracket[1, \sactivity]), (\transitionbracket[k, \eactivity], out)} \subseteq \arcs$.
\textbf{Adding and connecting course transitions:}
The set of activities is identical to the set of course IDs contained in the study plan, i.e., $\activities = dom(SP)$.
Finally, we add a transition~$\transition[cid]$ with $\labelingfunc(\transition[cid]) = cid$ for each $cid \in dom(SP)$ and connect them by their start and end term in the study plan, i.e., we add places $\aset{\placebracket[cid, in], \placebracket[cid, out] \mid cid \in dom(SP)} \subseteq \places$ and arcs $\aset{(\transitionbracket[{T_{start}}, \sactivity], \placebracket[cid, in]), (\placebracket[cid, in], \transition[cid]), (\transition[cid], \placebracket[cid, out]), \allowbreak(\placebracket[cid, out], \transitionbracket[{T_{end}}, \eactivity]) \mid cid \in dom(SP) \wedge (T_{start}, T_{end}) = SP(cid)} \subseteq \arcs$.
Additionally, we make the transition with the elective course ID $C_{el} \in dom(SP)$ repeatable, i.e., $\aset{(\placebracket[{C_{el}}, out], \transition[{C_{el}}]), (\transition[{C_{el}}], \placebracket[{C_{el}}, in])} \subseteq \arcs$, e.g., C01 in Figure \ref{fig:sp_petri_net_2010}.

We obtained two study plans $SP_{10}$ and $SP_{18}$ from the 2010 and 2018 study regulations for the RWTH Aachen University Computer Science B.Sc. program.
In Figure \ref{fig:sp_petri_net_2010} and Figure~\ref{fig:sp_petri_net_2018}, we show the workflow nets $\net[10]$ and $\net[18]$ constructed from the study plans $SP_{10}$ and $SP_{18}$, respectively.
We highlight transitions labeled with course IDs missing in either study plan or differing in start/end terms in both workflow nets, i.e., $\aset{\transition[cid] \mid (cid \in ((\difference{dom(SP_{10})}{dom(SP_{18})}) \cup (\difference{dom(SP_{18})}{dom(SP_{10}))}) \vee (cid \in (dom(SP_{10}) \cup dom(SP_{18})) \wedge SP_{10}(cid) \not = SP_{18}(cid)))}$.
The reserved elective course ID is C01 in both workflow nets.

\paragraph{Deriving Labeled Partial Orders.}
Next, we introduce how to derive labeled partial orders from an event log $\log$.
We derive two types of labeled partial orders: life-cycle-aware and atomic.
Life-cycle-aware modeling considers all exam attempts for a course ID, including failed attempts, while atomic modeling only considers the passed attempt.
This allows us to differentiate between students' intended conformance in exam-taking behavior and their actual conformance based only on passed attempts.

Let $\log \in \multisetset{\universeofevents}$ be an event log and $mat \in \universeofMatriculationNumbers$ be a student ID.
The filtered event log $\log[mat]$ of all exam attempts by a student with student ID $mat$ is the multiset $\log[mat] = \amulset{\event \in \log \mid \sel{Mat}(\event) = mat}$.
For each course ID $cid \in \aset{\sel{cid}(\event) \mid \event \in \log[mat]}$ in the student's event data, we calculate the start term $t_{\sactivity}(\log, mat, cid)$ and the end term $t_{\eactivity}(\log, mat, cid)$.
The end term is the passing term, i.e., $t_{\eactivity}(\log, mat, cid) = \sel{term}(\event)$ with $\event \in \log[mat] \wedge \sel{cid}(\event) = cid \wedge \sel{pass}(\event) = 1$.
The start term differs between life-cycle-aware and atomic modeling: atomic modeling uses a single point in time, i.e., $t_{\sactivity}(\log, mat, cid) = t_{\eactivity}(\log, mat, cid)$, whereas life-cycle-aware modeling uses the term of the first exam attempt, i.e., $t_{\sactivity}(\log, mat, cid) = min(\aset{\sel{term}(\event) \mid \event \in \log[mat] \wedge \sel{cid}(\event) = cid})$ with $min$ being the minimum function.

Let $\net = (\places, \transitions, \arcs, \activities, \labelingfunc)$ be a workflow net constructed from a study plan $SP$.
A labeled partial order $lpo_{mat} = (\partialNodes[mat], \partialRelation[mat], \activities[mat], \labelingfunc[mat])$ for a student ID $mat \in \universeofMatriculationNumbers$ that agrees with the workflow net $\net$ is derived as follows.
First, we ensure that the workflow net $\net$ and the labeled partial order $lpo$ agree by sharing the same activities, i.e., $\activities[mat] = \activities$.
Next, we add a node $\partialNode[mat, cid] \in \partialNodes$ for each course ID $cid \in \aset{\sel{cid}(\event) \mid \event \in \log \wedge \sel{mat}(\event) = mat}$ in the event log~$\log[mat]$.
During labeling, we differ between mandatory and elective courses, i.e., $\forall_{\partialNode[cid] \in \partialNodes} \colon \labelingfunc(\partialNode[mat, cid]) = cid$ if $cid \in (\difference{\activities}{\aset{C_{el}}})$ and $\labelingfunc(\partialNode[mat, cid]) = C_{el}$ otherwise.
Any two partial order nodes $\partialNode[mat, cid], \partialNode[mat, cid'] \in \partialNodes$ share an edge if one course's start term $t_{\sactivity}(\log, mat, cid')$ follows another course's end term $t_{\eactivity}(\log, mat, cid)$, i.e., $\forall_{\partialNode[mat, cid], \partialNode[mat, cid'] \in \partialNodes} \colon (t_{\eactivity}(\log, mat, cid) < t_{\sactivity}(\log, mat, cid')) \Rightarrow ((\partialNode[mat, cid], \partialNode[mat, cid']) \in\ \partialRelation)$.
Next, we define a function to calculate term distances between the courses in the event log and their expected times in the study plan.

\paragraph{Term Distance.}
Let $SP$ be a study plan, $\log$ an event log, $mat$ a student ID, and $cid$ a course ID.
For better readability, we conclude (1) $exp(\sactivity) = \sel{1}(SP(cid))$ if $cid \in dom(SP)$ and $exp(\sactivity) = \sel{1}(SP(C_{el}))$ otherwise, and (2) $exp(\eactivity) = \sel{2}(SP(cid))$ if $cid \in dom(SP)$ and $exp(\eactivity) = \sel{2}(SP(C_{el}))$ otherwise, the expected start and end term in the study plan, respectively.
The term $act = t_{\eactivity}(\log, mat, cid)$ is the actual term in which the student with the student ID $mat$ passed a course.
We define the distance function $\Delta_t(SP, \log, mat, cid)$ as follows:
$$
    \Delta_t(SP, \log, mat, cid) = 
    \begin{cases}
        \makebox[2.2cm]{0,\hfill} \text{if, } exp(\sactivity) \leq act \leq exp(\eactivity), &\makebox[1.7cm]{\hfill\text{(in time)}}\\
        \makebox[2.2cm]{act - exp(\sactivity),\hfill} \text{if, } act < exp(\sactivity), \text{and} &\makebox[1.7cm]{\hfill\text{(early)}}\\
        \makebox[2.2cm]{act - exp(\eactivity),\hfill} \text{if, } act > exp(\eactivity). &\makebox[1.7cm]{\hfill\text{(late)}}
    \end{cases}
$$

Thus, the distance function $\Delta_t(SP, \log, mat, cid)$ indicates if a course was taken on time, early, or late, showing the distance to the expected start or end term(s).
In our example, we know that all terms are semesters in the study plans and in the event log.
Thus, we simplify the distance function to a year distance $\Delta_y(SP, \log, mat, cid) = \left\lceil\frac{\Delta_t(SP, \log, mat, cid)}{2}\right\rceil$, summarizing two terms as up to one year.
Next, we align each course ID in the study plan and the labeled partial orders by their relative positions.
Both are combined later to derive insights.

\paragraph{Alignment Order Relation.}
Let $\alignment(lpo, \net)$ be a partial order alignment between a labeled partial order $lpo = (\partialNodes, \partialRelation, \activities', \labelingfunc')$ and a workflow net $\net = (\places, \transitions, \arcs, \activities, \labelingfunc)$.
By construction, all activities in the labeled partial order $lpo$ are contained in the workflow net $\net$, i.e., $\activities' \subseteq \activities$, and each workflow net activity occurs at least once on the model side $\alignmentModel$ of the alignment $\alignment$.
Thus, for each activity, or more precise course ID $cid \in \universeofCourseIDs$, on the log side $\alignmentLog$, we can determine its position relative to its position on the model side $\alignmentModel$ of the partial order alignment $\alignment$.

We distinguish four cases:
(1) A course ID occurs at the same position in the alignment and therefore the student's behavior and the study plan are aligned~(\Synchronous).
(2) A course ID occurs first on the log side then on the model side, indicating that the course is moved forward compared to the study plan~(\LogBeforeModel).
(3) A course ID occurs first on the model side and then on the log side, indicating that the course is moved back compared to the study plan~(\ModelBeforeLog).
(4) A missing entry on the log side~(\LogMissing).

Formally, for $i \in \range{1, \cardinality{\alignment}}$ a position in the alignment where the $i$-th entry on the model side is a course ID $cid \in \universeofCourseIDs$, i.e., $\alignmentModel(i) = cid$, we define:
\begin{itemize}
    \item \makebox[1.7cm]{\Synchronous,\hfill} if $\alignmentLog(i) = cid$, \hfill(Synchronous)
    \item \makebox[1.7cm]{\LogBeforeModel,\hfill} if $\exists_{j \in \range{1, i - 1}} \colon \alignmentLog(j) = cid$, \hfill (Student before Plan)
    \item \makebox[1.7cm]{\ModelBeforeLog,\hfill} if $\exists_{j \in \range{i + 1, \cardinality{\alignment}}} \colon \alignmentLog(j) = cid$, and \hfill (Student after Plan)
    \item \makebox[1.7cm]{\LogMissing,\hfill} if $\forall_{j \in \range{1, \cardinality{\alignment}}} \colon \alignmentLog(j) \not = cid$. \hfill (Missing in Log)
\end{itemize}

Next, we apply our approach to a real-life data set to demonstrate its applicability and gain insights.
\section{Evaluation}\label{sec:eval}
We conducted the evaluation using PM4Py \footnote{https://github.com/pm4py/pm4py-core}, an open-source Python library for process mining.
For evaluation, we used a real-life data from three German universities collaborating in the BMBF-funded joint research project ``AIStudyBuddy''.
As part of the project, two software tools are being developed to assist students in organizing their studies and study planners in analyzing study programs.
The data includes 1,162 RWTH Aachen University Computer Science Bachelor's graduates who began earliest in 2010 and have at least 150 ECTS.
As mentioned in Section \ref{sec:approach}, in Figure \ref{fig:sp_petri_net_2010} and Figure \ref{fig:sp_petri_net_2018} the corresponding plans for 2010 and 2018 are modeled as workflow nets $\net[10]$ and $\net[18]$.
For each student from the event log, a labeled partial order is created using the life-cycle-aware and atomic approach.
The code for modeling study plans and translating educational event logs is publicly available on GitLab\footnote{\label{note1}\url{https://git.rwth-aachen.de/christian.rennert/po-based-SP-eval}}.
In addition, the published code includes mock student data and the two study plans used in this work.
We computed alignments for both workflow nets using both, the atomic or life-cycle-aware labeled partial orders, and selected the study plan with the lower alignment cost for further result aggregation.
We determined the expected study plan based on whether a student started before or from 2018 onward.
Table~\ref{tab:expected-most-conforming-heatmap} shows a confusion matrix for the expected and the most conforming study plan.

\noindent
\begin{minipage}{0.62\linewidth}
    \strut
    \-\hspace{1em}
    Of the 860 students, 237, or around $27.5\%$, expected to follow the 2010 study plan conform most to the 2018 plan.
    In contrast, only a fraction of around $13.2\%$ of those expected to follow the 2018 plan conform most to the 2010 plan.
    An explanation may be that in an educational setting, students may adopt to a new study plan within their studies.
    It is to note, that the confusion matrix does not alter between the life-cycle-aware and atomic partial order modeling approach.
\end{minipage}\hfill
\begin{minipage}{0.36\linewidth}
%\resizebox{\textwidth}{!}{
    \begin{minipage}{\linewidth}
        \centering
        \begin{tabular}{ccccc}
                                                            & \multicolumn{4}{c}{Most conforming}                                                              \\
        \multirow{4}{*}{\rotatebox[origin=c]{90}{Expected}} & \multicolumn{1}{c|}{}         & \multicolumn{1}{c|}{2010} & \multicolumn{1}{c|}{2018} & $\Sigma$ \\ \cline{2-5} 
                                                            & \multicolumn{1}{c|}{2010}     & \multicolumn{1}{c|}{623}  & \multicolumn{1}{c|}{237}  & 860      \\ \cline{2-5} 
                                                            & \multicolumn{1}{c|}{2018}     & \multicolumn{1}{c|}{40}   & \multicolumn{1}{c|}{262}  & 302      \\ \cline{2-5} 
                                                            & \multicolumn{1}{c|}{$\Sigma$} & \multicolumn{1}{c|}{663}  & \multicolumn{1}{c|}{499}  & 1162    
        \end{tabular}
        \captionsetup[table]{font=footnotesize}
        \captionof{table}{Confusion matrix between expected study plans and most conforming study plans.}
        \label{tab:expected-most-conforming-heatmap}
    \end{minipage}
%}
\end{minipage}

\begin{table}[t]
    \caption{\smaller Distance in years for the 2010 study plan and life-cycle-aware student courses.}
    \label{tab:2010}
\resizebox{\textwidth}{!}{
\begin{tabular}{ccccccccccccccccccc|c}
\toprule
                       & \multicolumn{6}{c}{Student before Plan (\LogBeforeModel)}                                                                                              & \multicolumn{6}{c}{Synchronous (\Synchronous)}                                                                                                        & \multicolumn{6}{c|}{Student after Plan (\ModelBeforeLog)}                                                                                             & \multirow{2}{*}{\begin{tabular}[c]{@{}c@{}}Missing in\\Log (\LogMissing)\end{tabular}} \\ \cline{1-19}
$\Delta$ years               & \makebox[0.6cm]{$\leq$-2} & \makebox[0.6cm]{-1} & \makebox[0.6cm]{0} & \makebox[0.6cm]{1} & \makebox[0.6cm]{2} & \makebox[0.6cm]{$\geq3$}  & \makebox[0.6cm]{$\leq$-2} & \makebox[0.6cm]{-1} & \makebox[0.6cm]{0} & \makebox[0.6cm]{1} & \makebox[0.6cm]{2} & \makebox[0.6cm]{$\geq3$}    & \makebox[0.6cm]{$\leq$-2} & \makebox[0.6cm]{-1} & \makebox[0.6cm]{0} & \makebox[0.6cm]{1} & \makebox[0.6cm]{2} & \makebox[0.6cm]{$\geq3$} &                                                                           \\ \midrule
\multicolumn{1}{c|}{C05} & 0 & 9 & 29 & 1 & 1 & \multicolumn{1}{c|}{0} &1 & 2 & 570 & 27 & 5 & \multicolumn{1}{c|}{2} & 0 & 0 & 0 & 11 & 3 & 2 & 0\\
\multicolumn{1}{c|}{C12} & 13 & 0 & 1 & 0 & 0 & \multicolumn{1}{c|}{0} &2 & 16 & 493 & 100 & 25 & \multicolumn{1}{c|}{6} & 0 & 0 & 3 & 4 & 0 & 0 & 0\\
\multicolumn{1}{c|}{C13} & 0 & 2 & 4 & 5 & 0 & \multicolumn{1}{c|}{0} &2 & 20 & 428 & 138 & 34 & \multicolumn{1}{c|}{20} & 0 & 0 & 0 & 0 & 0 & 0 & 10\\
\multicolumn{1}{c|}{C16} & 0 & 0 & 0 & 0 & 0 & \multicolumn{1}{c|}{0} &4 & 7 & 530 & 13 & 3 & \multicolumn{1}{c|}{1} & 0 & 0 & 0 & 60 & 22 & 23 & 0\\
\multicolumn{1}{c|}{C17} & 77 & 21 & 3 & 2 & 0 & \multicolumn{1}{c|}{1} &0 & 119 & 111 & 65 & 10 & \multicolumn{1}{c|}{4} & 0 & 0 & 0 & 0 & 0 & 0 & 250\\
\multicolumn{1}{c|}{C18} & 0 & 3 & 3 & 0 & 1 & \multicolumn{1}{c|}{0} &1 & 2 & 456 & 21 & 7 & \multicolumn{1}{c|}{1} & 0 & 0 & 1 & 69 & 57 & 41 & 0\\ \bottomrule
\end{tabular}}

    \caption{\smaller Distance in years for the 2018 study plan and life-cycle-aware student courses.}
    \label{tab:2018}
\resizebox{\textwidth}{!}{
\begin{tabular}{ccccccccccccccccccc|c}
\toprule
                       & \multicolumn{6}{c}{Student before Plan (\LogBeforeModel)}                                                                                              & \multicolumn{6}{c}{Synchronous (\Synchronous)}                                                                                                        & \multicolumn{6}{c|}{Student after Plan (\ModelBeforeLog)}                                                                                             & \multirow{2}{*}{\begin{tabular}[c]{@{}c@{}}Missing in\\Log (\LogMissing)\end{tabular}} \\ \cline{1-19}
$\Delta$ years               & \makebox[0.6cm]{$\leq$-2} & \makebox[0.6cm]{-1} & \makebox[0.6cm]{0} & \makebox[0.6cm]{1} & \makebox[0.6cm]{2} & \makebox[0.6cm]{$\geq3$}  & \makebox[0.6cm]{$\leq$-2} & \makebox[0.6cm]{-1} & \makebox[0.6cm]{0} & \makebox[0.6cm]{1} & \makebox[0.6cm]{2} & \makebox[0.6cm]{$\geq3$}    & \makebox[0.6cm]{$\leq$-2} & \makebox[0.6cm]{-1} & \makebox[0.6cm]{0} & \makebox[0.6cm]{1} & \makebox[0.6cm]{2} & \makebox[0.6cm]{$\geq3$} &                                                                           \\ \midrule
\multicolumn{1}{c|}{C05} & 2 & 3 & 25 & 2 & 1 & \multicolumn{1}{c|}{0} &1 & 6 & 371 & 46 & 9 & \multicolumn{1}{c|}{1} & 0 & 0 & 3 & 19 & 6 & 4 & 0\\
\multicolumn{1}{c|}{C12} & 7 & 54 & 32 & 8 & 7 & \multicolumn{1}{c|}{2} &1 & 52 & 260 & 53 & 16 & \multicolumn{1}{c|}{7} & 0 & 0 & 0 & 0 & 0 & 0 & 0\\
\multicolumn{1}{c|}{C13} & 0 & 0 & 0 & 0 & 0 & \multicolumn{1}{c|}{0} &0 & 1 & 250 & 160 & 51 & \multicolumn{1}{c|}{33} & 0 & 0 & 0 & 0 & 0 & 0 & 4\\
\multicolumn{1}{c|}{C16} & 0 & 0 & 0 & 0 & 0 & \multicolumn{1}{c|}{0} &6 & 5 & 326 & 16 & 2 & \multicolumn{1}{c|}{0} & 0 & 0 & 1 & 75 & 44 & 24 & 0\\
\multicolumn{1}{c|}{C17} & 3 & 3 & 2 & 0 & 0 & \multicolumn{1}{c|}{0} &1 & 14 & 262 & 50 & 13 & \multicolumn{1}{c|}{11} & 0 & 0 & 0 & 0 & 0 & 0 & 140\\
\multicolumn{1}{c|}{C18} & 0 & 4 & 3 & 1 & 0 & \multicolumn{1}{c|}{0} &0 & 6 & 282 & 37 & 5 & \multicolumn{1}{c|}{4} & 0 & 1 & 1 & 87 & 30 & 38 & 0\\ \bottomrule
\end{tabular}}
\end{table}

The alignment of the most conforming study plan is used to compute the alignment order relation and year distance for each course.
We count the frequency of each combination for the two study plans.
Tables \ref{tab:2010} and \ref{tab:2018} show the results for both study plans, respectively.
Complete and further tables are available on Gitlab\footnote{See Footnote \ref{note1}.}, showing all distances in terms and years and all courses being included as both life-cycle-aware and atomic.
Although present, the difference between atomic and life-cycle-based modeling is marginal in our results.
However, we expect that this may vary when future experiments are applied, and therefore requires further investigation.
In the following, we focus only on life-cycle-aware modeling.

Next, we analyze the results to derive some potentially interesting insights for study planners. 
We select representatives from two groups of courses represented in both the 2010 and 2018 study plans:
(1) Courses moved in the 2018 study plan compared to 2010, i.e., courses with course IDs C12, C13, and C17, and
(2) courses not moved between the 2010 and 2018 study plans, i.e., courses with course IDs C05, C16, and C18.

For the former, we use our method to check if the changes are supported from results of the original 2010 study plan and if conformance improves. 
Therefore, we detail every change, describe the evidence, and evaluate the change in conformance.
In the 2010 study plan, Course C12 is taken in the fourth to fifth term, but in 2018, it shifts to the fifth to sixth term.
We do not identify direct results supporting this change.
After the change, instances of students taking a course earlier than its expected position increased from 14 to 110.
This holds even in relative counts, with more students conforming to the 2010 study plan than to the 2018 study plan.
This suggests the change may need more time to take effect or may be impractical for students.
For Course C17, the opposite holds.
In the 2010 plan, it is expected in the sixth term only, but in the 2018 plan, in the second to fifth term.
There is evidence of students taking the course earlier than planned and typically over 2 years ahead.
Additionally, more events are missing in the logs for the 2010 study plan than for the 2018 study plan, strengthening the finding.
Thus, we expect the change to positively impact conformance, as shown in the 2018 study plan results.
Course C13 moves from the fifth and sixth terms to the third to fifth terms.
There is minimal evidence of students taking the course early before the change and none after.
In conclusion, we expect our approach to be usable for detecting and evaluating potential changes effectively using the available evidence.

For the second group, we investigate the general misalignment quota and its stability after other course changes.
The misalignment quota may assess the course conformance as shown in the following.
Courses C16 and C18, preliminary courses in the first and third term, are often delayed due to their difficulty.
Life-cycle-aware course modeling together with ``student after plan'' counts shows individual delays of first exam attempts, i.e., for course C16, 75 students postpone by 1 year, 44 by two years, and 24 by three years.
This information is relevant to study planners because preliminary courses contain essential knowledge for other courses and may lead to a longer study duration, which planners aim to avoid.
We expect this method to provide study planners with a measurement of the frequency and role of delayed courses, serving as a basis for countermeasures like improved mentoring.
Toward stability, the relative frequency of students taking courses C16 and C18 has increased, from 105 out of 558 to 143 out of 355 students.
This may highlight the increasing importance of individual study planners' countermeasures and how to find them.

\paragraph{Future Work.}
We plan to apply the approach to more study plans and event data to further investigate the insights from the proposed method.
In addition, an adaptation is to cluster students by key performance indices such as GPA or study duration to explore their relation to (non)conforming behavior.
Finally, we plan to add the relative alignment positions to the event data to identify frequent itemsets of mismatched courses.
\section{Conclusion}\label{sec:conclusion}
This work contributes to the field of educational process mining by combining conformance checking methods to account for the relation between course order and the difference between expected and actual course passing terms.
We model study plans with workflow nets and student course-taking behavior with labeled partial orders, providing a basis for further experiments in educational process mining.
Partial order alignments are used to compare expected and actual course orders.
Using real-life data, initial experiments show our contribution's usability and may suggest valuable insights for study planners.

\phantom{}

\begin{credits}
\noindent
\begin{minipage}[c]{0.749\linewidth}
    \subsubsection{\ackname} We thank the Alexander von Humboldt (AvH) Stiftung for supporting our research (grant no. 1191945).
    The authors gratefully acknowledge the financial support by the Federal Ministry of Education and Research (BMBF) for the joint project AIStudyBuddy (grant no. 16DHBKI016).
    Further, we thank all reviewers for their valuable feedback
\end{minipage}\hfill
\begin{minipage}[c]{0.25\linewidth}
    \includegraphics[scale=0.4]{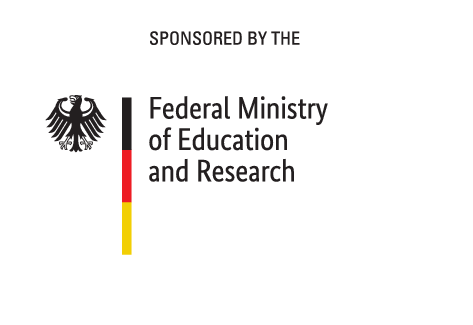}
\end{minipage}
\end{credits}

\bibliographystyle{splncs04}
\bibliography{mybibliography}
\end{document}